\documentclass[nonacm]{acmart}

\AtBeginDocument{%
  }

\usepackage{rotating} 
\usepackage{tabularx}
\usepackage{longtable}
\usepackage{subcaption}
\providecommand{\noopsort}[1]{}

\begin{document}

\title{Virtual Reality in Service Design for Plastics Recycling: Two Application Cases}

\author{Ashley Colley}
\orcid{0009-0008-5967-2828}
\affiliation{%
  \institution{University of Lapland}
  \city{Rovaniemi}
  \country{Finland}
}
\email{ashley.colley@ulapland.fi}

\author{Kuisma Hurtig}
\orcid{0009-0003-0375-9725}
\affiliation{%
  \institution{University of Lapland}
  \city{Rovaniemi}
  \postcode{96300}
  \country{Finland}}
\email{kuisma.hurtig@ulapland.fi}

\author{Juri Etto}
\orcid{0009-0005-4415-5906}
\affiliation{%
  \institution{University of Lapland}
  \city{Rovaniemi}
  \postcode{96300}
  \country{Finland}}
\email{juri.etto@ulapland.fi}

\author{Emma Kirjavainen}
\orcid{0000-0002-5181-6865}
\affiliation{%
  \institution{University of Lapland}
  \city{Rovaniemi}
  \postcode{96300}
  \country{Finland}}
\email{emma.kirjavainen@ulapland.fi}

\author{Pavlo Ivanov}
\affiliation{%
  \institution{University of Lapland}
  \city{Rovaniemi}
  \postcode{96300}
  \country{Finland}}
\email{pavlo.ivanov@ulapland.fi}

\author{Jonna Häkkilä}
\orcid{0000-0003-2172-6233}
\affiliation{%
  \institution{University of Lapland}
  \city{Rovaniemi}
  \country{Finland}
}
\email{jonna.hakkila@ulapland.fi}

\renewcommand{\shortauthors}{Colley et al.}

\begin{abstract}
Plastics recycling depends on everyday sorting practices and on how recycling services are communicated and experienced. Virtual reality (VR) can present these practices and services in situated, interactive form, yet its role in service design for plastics recycling is still emerging. This paper examines how VR tools can contribute to designing plastics recycling services through two application cases that address different stages of the recycling journey. The first case, Clean Cabin Escape, is a household scale VR escape room where players collect and sort waste items into locally relevant categories, with immediate feedback that supports practice with plastics recycling decisions. The second case is a VR simulation of a plastics recycling center that represents a real planned site and is used in service design workshops where stakeholders explore layout, signage and customer paths for plastics fractions. Across the cases, we analyse how VR supported learning, engagement and shared sensemaking, and how it interacted with other service design methods such as workshops, customer path mapping and physical artefacts. The findings show that VR can make domestic sorting tasks and complex recycling centers more concrete for both citizens and professionals, but also highlight trade offs related to hardware access, onboarding effort, visual fidelity and localisation of recycling rules. The paper concludes by outlining opportunities for integrating VR into broader service design toolsets for plastics recycling and circular economy services, and by pointing to directions for future research on long term impact and inclusive design.
\end{abstract}


\maketitle

\section{Introduction}

Plastic waste is a persistent environmental challenge, with packaging and everyday consumer products contributing heavily to municipal waste streams. Recycling is a key strategy for reducing environmental impact, yet recycling rates remain below policy targets in many regions. People often struggle with sorting rules, multi-material packaging, and local variations in recycling guidelines, which leads to contamination of waste streams and missed opportunities for material recovery \cite{vogt2021recycling,chen2022current}. This situation calls for approaches that do more than inform citizens about rules but instead engage them in practical, situated learning about recycling.

Human–computer interaction (HCI) research has a long-standing interest in sustainability and in supporting more sustainable everyday practices through interactive technologies \cite{blevis2017further,karhu2019collaborative,mencarini2024imagining}. In this context, gamification and virtual reality (VR) offer ways to create interactive experiences that combine learning with challenge and enjoyment \cite{lampropoulos2024virtual,Novo2024-fx}. VR, in particular, makes it possible to rehearse real-world tasks in controlled, immersive environments, where users can experiment, make mistakes, and receive immediate feedback on their actions \cite{gong2022systematic}. This combination is promising for recycling education, where correct behaviour depends on repeated practice and confidence in handling ambiguous items.

At the same time, plastics recycling is not only a matter of individual knowledge, but also of how recycling services are organised and experienced. Service design approaches conceptualise recycling systems as service journeys that include learning local rules, organising waste at home, visiting collection points or recycling centers, and interacting with physical and informational touchpoints \cite{miettinen2017introduction,stickdorn2018service}. In this view, recycling centers and household waste points are part of a broader servicescape \cite{bitner1992servicescapes} that shapes how understandable, welcoming, and actionable plastics recycling appears to citizens. VR has started to be used as a design tool in such settings, e.g., in virtual bodystorming and service walkthroughs that allow designers and stakeholders to experience future services before implementation \cite{boletsis2017virtual,boletsis2018virtual,colley2020exploring}.

This paper connects these strands by examining how VR tools can support plastics recycling both as an educational topic and as a service design challenge. We focus on the question of \emph{how VR tools can be used as part of a toolset for the service design of plastics recycling}. We approach this question through two application cases that address different stages and scales of the recycling journey. The first case, \emph{Clean Cabin Escape}, is a VR escape room game set in a log cabin, where players must collect and sort scattered waste items into the correct recycling bins. The game integrates waste-sorting rules with familiar domestic actions and uses game mechanics and feedback to encourage repeated practice \cite{etto2025gamified}. The second case presents a VR simulation of a plastics recycling center, developed as part of a broader service design process for a real recycling site, where users navigate the yard, locate the plastics area, and sort items using multi-level signage while designers and stakeholders use the environment to discuss, prototype, and refine the service concept \cite{etto2024designing}.

This article extends the contribution of prior demonstrations of each of the cases \cite{etto2025gamified,etto2024designing} by offering a combined analysis of the two cases through a service design perspective, bringing together their design rationales, empirical findings, and cross-case implications. The contribution of the paper is threefold. First, we describe the design and implementation of the two VR applications and their roles in plastics recycling. Second, we discuss how VR fits alongside other service design methods, such as workshops, physical prototypes, and visualised service journeys, when working on plastics recycling services. Third, we reflect on opportunities and limitations for integrating VR tools into a service design toolset for plastics recycling, outlining directions for future work on sustainable service design.

\section{Related Work}

This section situates our work at the intersection of recycling education, game-based learning, and service design. We first look at studies on plastics recycling and citizen engagement, then turn to gamification, escape rooms, and VR in learning contexts, and finally discuss work that links service design with VR as a design tool.

\subsection{Recycling education and plastics recycling challenges}

This subsection focuses on research that examines how people learn recycling practices and how plastics recycling is communicated in everyday life. The aim is to outline key barriers in citizen engagement and highlight prior interventions that inform our VR-based approach.

Recycling education is a key element in discussions of global environmental issues and the move toward circular economy systems. Prior research has examined how schools, universities and public campaigns can influence knowledge and everyday habits related to waste sorting and plastics use \cite{bennett2021informing,Sidian2024-dy,Soegoto2021-cr,Cheng2021-ny,Sahroni2022-po,geissdoerfer2017circular}. These works highlight confusion about material categories, limited feedback on correct sorting and the importance of guidance that links rules to concrete household practices \cite{bennett2021informing,Sidian2024-dy,lidia2018encourage,ro2017making}. Studies on plastics recycling interventions report a range of tools, including mobile applications, interactive installations, and educational activities that seek to encourage sorting and describe circular economy processes in more tangible ways \cite{Novo2024-fx,Bai2024-nc,Mei2019-ww,venturi2025trash,santti2020digitalization,aguiar2019gamification}. 

Gamification approaches applied to environmental topics frequently draw on earning points, progress indicators, and other game elements to influence behaviour. Work on recycling support systems and eco-feedback applications suggests that such elements can increase engagement in the short term and provide structure for practising desired actions \cite{Novo2024-fx,gaggi2020learning,venturi2025trash,Cheng2021-ny,Siregar_undated-od}. Palmquist's ethnographic study of a gamified recycling house argues that playful formats can deepen citizens' grasp of circular economy stages rather than serve only as simple reward systems \cite{palmquist2022design}. These insights motivate the use of game-based approaches for plastics recycling, particularly in settings where people face complex sorting rules and multi-step processes.

\subsection{Gamification, escape rooms and VR for learning}

Here, we consider game-based and VR-based learning approaches that relate to our two application cases. The focus is on how escape rooms, gamification, and immersive environments have been used to support engagement, collaboration, and procedural learning.

Escape room activities form one family of game-based learning approaches examined in environmental and sustainability education. Prior work reports that time pressure, puzzles and narrative framing can support motivation and collaboration and give participants a memorable context for key concepts and procedures \cite{Bai2024-nc,Mei2019-ww,venturi2025trash}. Studies on digital and hybrid escape rooms extend these ideas to settings where players interact with virtual environments, digital locks or mixed physical-digital artefacts \cite{Novo2024-fx,arbesser2024project,sangeetha2022leveraging}. 

VR-based learning experiences extend this line of research. Reviews and empirical studies report effects on immersion, engagement and learning outcomes across domains including science education, professional training and empathy building \cite{lampropoulos2024virtual,gong2022systematic,alqallaf2024empathy,plotzky2021virtual,kaminska2017virtual}. In environmental and energy-related contexts, VR has been used to stage scenarios about resource use, climate impacts and waste management, sometimes combined with game elements or interactive tasks \cite{Novo2024-fx,Bai2024-nc,Mei2019-ww,venturi2025trash,pham2018energy}. Research on VR and AR for procedural training and safety-related tasks shows how virtual environments can support repeated practice of multi-step activities and the handling of potentially hazardous situations \cite{salah2019virtual,samaddar2024ar}. These findings inform the design of VR experiences where users identify items, apply sorting rules and receive feedback on their actions in a simulated setting.

\subsection{Service design and VR as a design tool}

This subsection turns to service design research and the use of VR as a design material in service projects. The goal is to show how service design concepts, methods and VR-based prototypes frame our work on plastics recycling services.

Service design literature describes services as systems of interactions, touchpoints and backstage processes. Introductory texts and handbooks highlight co-creation, visualisation and iterative experimentation as central ideas in this field \cite{miettinen2017introduction,stickdorn2018service,magnusson2003managing}. Customer path mapping and service blueprints make user actions and contact points visible for design work and connect with HCI discussions of user experience across channels and over time \cite{folstad2018customer,yap2021hci,bitner1992servicescapes}. Service design draws on industrial and interaction design traditions and uses a broad repertoire of methods. Curedale describes method sets that include stakeholder mapping, personas, path maps and co-design workshops familiar from user-centred design \cite{curedale2013service,miettinen2017introduction}. Prototypes play a central role, ranging from enactments and storyboards to mock-ups of digital and physical touchpoints that support communication and iteration with stakeholders \cite{bae2014visual,holmlid2007prototyping,razek2019comparing}. 

Research that links service design and digital technology treats technology both as an object of design and as a design material. Interaction-oriented perspectives help to make sense of how services unfold over time and across channels, and later work has described multi-level views of service systems and digital service paths \cite{holmlid2007interaction,patricio2018upframing,fallman2008interaction,alhonsuo2022digital}. Recent work connects service design with extended reality tools and proposes frameworks for VR-supported service design processes in commercial and public-sector projects \cite{hakkila2023service,roto2021overlaps}. 

Studies that report concrete uses of VR in service design describe a variety of contexts. Boletsis et al. explored VR simulations of public spaces, including duty-free shops and waiting areas, to support discussion of layout, signage and customer experience with stakeholders \cite{boletsis2017virtual,boletsis2018virtual,kwon2015evaluating}. In museum and exhibition projects, VR has been used to represent spaces, actors and artifacts and to rehearse visitor paths before physical installations are finalised \cite{pierroux2021exhibitions,hafner2016virtual}. Comparative studies examine differences between large-screen projection, VR and AR as media for presenting future service situations in workshops, reporting that VR can support focus on the environment but may constrain non-verbal communication between participants \cite{colley2020exploring,bae2014visual,voit2019online}. Work on virtual fieldwork extends these ideas to outdoor and open settings where walking, distances, and spatial relationships are part of the service experience \cite{gushima2021virtual,boletsis2017virtual}. 

Within services for circular economy and waste management, research has examined how service concepts, infrastructures and communication strategies can promote plastics recycling and other pro-environmental practices \cite{geissdoerfer2017circular,lidia2018encourage,ro2017making,palmquist2022design}. Interactive technologies appear in this space as both touchpoints and design tools, e.g., in visualisations, digital platforms and VR-based prototypes \cite{Novo2024-fx,venturi2025trash,roto2021overlaps,plotzky2021virtual,kaminska2017virtual}. Virtual environments that mirror real or planned service settings provide one more option in the service designer's toolbox, since they can replicate locations, procedures and multi-step interactions at relatively low marginal cost once they exist \cite{boletsis2017virtual,boletsis2018virtual,bae2014visual,plotzky2021virtual}. The VR applications studied in this paper build on this prior work by combining game-based recycling education with VR-supported service design of a plastics recycling center.

\section{Case 1: Clean Cabin Escape - VR Escape Room for Household Plastics Recycling}

\subsection{Case context and goals}

\begin{figure}
    \centering
    \includegraphics[width=1\linewidth]{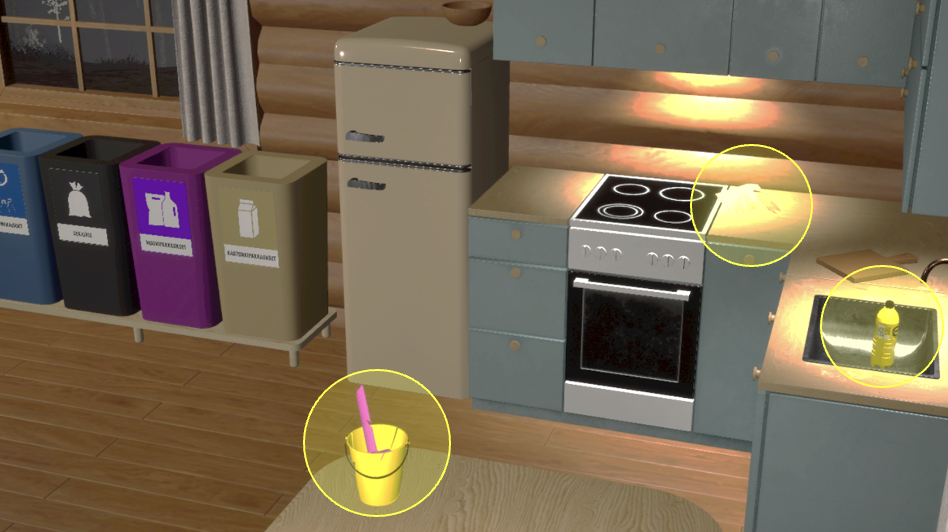}
    \caption{The interior of the cottage highlighting some of the plastic items to be collected and recycled to complete the escaperoom game.}
    \label{fig:cottage_items}
\end{figure}

\begin{figure}
    \centering
    \includegraphics[width=1\linewidth]{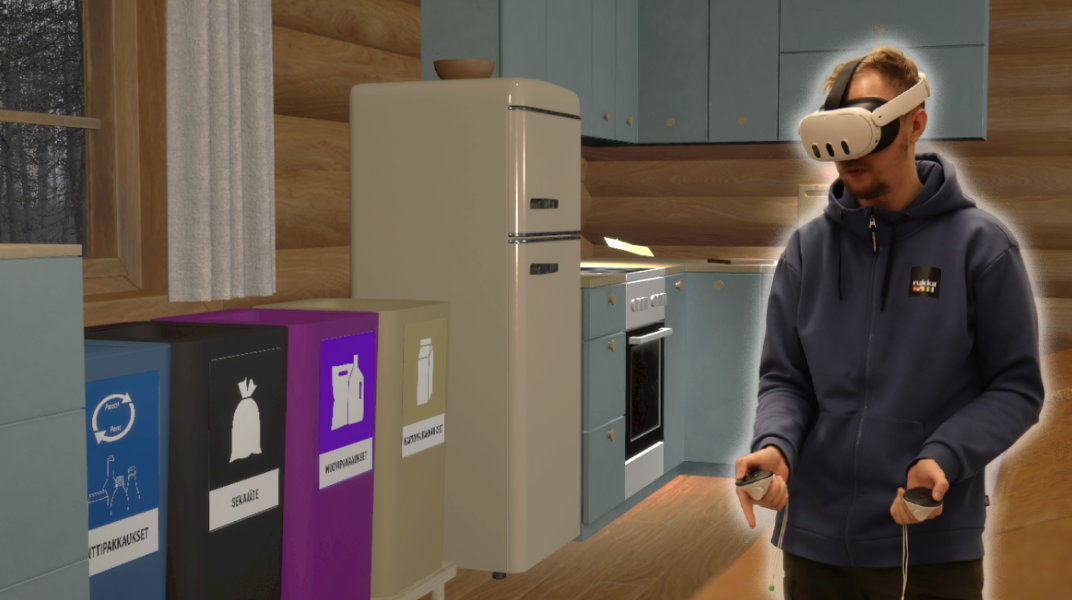}
    \caption{Mock up image illustrating the user interaction with the VR escape game.}
    \label{fig:cottage_mockup}
\end{figure}

The \textit{Clean Cabin Escape} game is a virtual reality escape room set in a log cabin in a forest. The setting draws on familiar Nordic cottage culture and everyday waste practices, placing players in a scenario where correct plastics recycling becomes the key to escaping the cabin \cite{etto2025gamified}. The game was created as a prototype for recycling education, combining immersive VR, escape room mechanics and local recycling rules. The aim is to help players rehearse sorting decisions for plastic and other household waste, increase confidence in handling ambiguous items and present recycling as an engaging problem-solving activity rather than a chore.

The project addresses two related goals. First, it explores how VR escape rooms can support learning about plastics recycling and waste sorting in a way that links abstract rules to situated actions in a domestic environment. Second, it probes how the format could fit into a broader service portfolio for organisations that provide recycling education or manage waste services, e.g., in schools, youth activities, museums or public outreach events.

\subsection{Game and interaction design}

The \textit{Clean Cabin Escape} environment was modelled as a compact log cabin interior with tables, sofas and storage surfaces populated by everyday waste items (Figure \ref{fig:cottage_items}). The virtual environment and game objects were created using Blender and Substance Painter, with some items modelled via 3D scanning of real objects. These assets were imported into Unreal Engine, where the game logic and interactions were implemented, with gameplay running on a standalone VR headset such as Meta Quest 3.

At the start of the experience, the player finds themselves locked inside the cabin and must complete a recycling challenge to open the door. Various pieces of trash are scattered across the room. The core task is to locate these items, pick them up using VR controller interactions and sort them into four recycling bins that match local categories: mixed waste, plastic packaging, cardboard packaging and returnable bottles with a deposit (Figure \ref{fig:cottage_mockup}). Correct sorting gradually progresses the game state toward escape.

The game emphasises clear feedback and repetition of correct actions. When a player places an item in a bin, colored lights on the lid indicate whether the choice is correct or incorrect. Items that are mis-sorted reappear near the bins, decreasing the effort needed for repeated attempts and keeping attention on the sorting decision rather than navigation. This supports trial-and-error learning of ambiguous categories such as composite packaging. The number and variety of items are tuned to keep the experience compact while still covering a representative set of plastics and related waste fractions.

Escape room conventions structure the experience around a single overarching puzzle: “clean the cabin correctly to get out”. Time pressure is moderate to avoid stress, and retain the focus on exploration and learning. The visual style aims to balance realism with a slightly stylised aesthetic to keep the environment legible and comfortable for a broad audience.

\subsection{Evaluation}

An initial evaluation of \textit{Clean Cabin Escape} was conducted with a sample of 11 participants representing a range of ages and backgrounds relevant for recycling education. The evaluation aimed to probe three areas: perceived engagement, perceived learning about recycling and practical usability of the VR interaction. Participants played the game individually in a VR lab setting. They were encouraged to think aloud during play while an observer took notes on their behaviour, difficulties and comments about the environment, items and bins. After the session, a short questionnaire captured ratings of enjoyment, perceived usefulness for learning about recycling and interest in playing similar games in the future. Open-ended questions invited comments on what felt clear or confusing about the sorting tasks and the VR interaction.

Participants reported high enjoyment and strong interest in the game format as a way to learn about recycling. Many commented that the cabin environment felt inviting and that the scattered items created a sense of “mess” that they wanted to clean up. Participants indicated that the four-bin system and light-based feedback helped them understand or reinforce distinctions between mixed waste, plastic packaging, cardboard and returnable bottles, especially for items that might otherwise have been ambiguous. Several reported that the VR setting made them more focused on the task than they would be with printed materials or a website.

At the same time, the evaluation surfaced practical issues in interaction and comfort. Some participants needed guidance on VR locomotion and grabbing mechanics, and a few experienced mild motion discomfort. The study also highlighted the time required for onboarding new VR users, which is relevant if the game is deployed in schools or public events.

\subsection{Findings and interaction design challenges}

The evaluation suggests that immersive VR escape rooms can provide a memorable, engaging frame for learning household recycling practices, echoing findings from VR and game-based sustainability education \cite{Novo2024-fx,Bai2024-nc,Mei2019-ww,venturi2025trash,pham2018energy,salah2019virtual,arbesser2024project}. The cabin setting anchors recycling in a domestic context, while the escape goal and scattered items give concrete reasons to pick up and classify waste. Participants’ comments indicate that the combination of spatial exploration, manipulation of 3D objects and immediate visual feedback supports reflection on sorting rules, in line with broader findings on immersive VR for behaviour adoption and environmental topics \cite{lampropoulos2024virtual,taufik2021changing,alqallaf2024empathy}.

The study also highlights several interaction design challenges. Getting VR novices comfortable with headset use, controller mappings and locomotion adds overhead, particularly for short educational sessions. This resonates with reports that VR learning experiences must balance immersion with accessibility and physical comfort \cite{plotzky2021virtual,kaminska2017virtual}. In our case, small adjustments such as simplifying movement controls, reducing unnecessary head motion and providing clear onboarding instructions emerged as priorities for future iterations.

The design of feedback presents another compromise. Strong negative signals for incorrect sorting, such as bright red lights or loud sounds, can make errors salient but risk discouraging experimentation if they feel punishing. At the same time, overly subtle feedback may fail to support learning. The current design adopts visual feedback that signals correctness without harsh penalties, paired with reappearance of mis-sorted items near the bins, but further work is needed to fine-tune these elements. The evaluation raised questions about scalability and access. VR headset cost and availability currently limit large-scale deployment, even though VR facilities in schools and libraries are becoming more common. This aligns with broader discussions of VR in education, where promising learning effects coexist with concerns about infrastructure and staffing \cite{salah2019virtual,lampropoulos2024virtual}. 

Potential future work includes refining the game mechanics, e.g., by experimenting with time limits, limited lives, scoring systems or alternative forms of feedback for correct and incorrect actions, and expanding the item set to cover a wider range of plastics and packaging formats. The authors also propose extending the concept to other sustainability themes beyond plastics recycling, such as energy conservation or food waste, using the escape room structure as a flexible template \cite{Novo2024-fx,Bai2024-nc,venturi2025trash}.

\subsection{Implications for plastics recycling service design}

The \textit{Clean Cabin Escape} provides one example of how VR can fit into a broader toolset for plastics recycling education. The game focuses on the household segment of the recycling journey, where citizens handle packaging and everyday waste, and connects directly to bin categories and visual cues used in local systems. In a service design context, the escape room can act as a touchpoint in educational or promotional services run by municipalities, waste management companies, NGOs or schools. 

From a Service Design perspective, the game functions both as part of the service offering and as a prototype for discussions with stakeholders. It can be used in workshops where participants first play the game and then reflect on their experiences, compare in-game bins and signs with real-world ones, and suggest improvements to communication materials or physical infrastructure. Insights from play sessions, such as which items confuse or which feedback patterns are most helpful, can feed into other parts of the service design process, including signage redesign, information campaigns and complementary non-VR tools. In this way, the cabin escape room case illustrates how an immersive, playful VR experience can complement more traditional service design methods for plastics recycling, by offering a simulated domestic setting where sorting practices and communication choices can be explored in action before or alongside real-life changes.

\section{Case 2: VR Simulation of a Plastics Recycling Center for Service Design}

\subsection{Service context and design brief}

\begin{figure}
    \centering
    \includegraphics[width=1\linewidth]{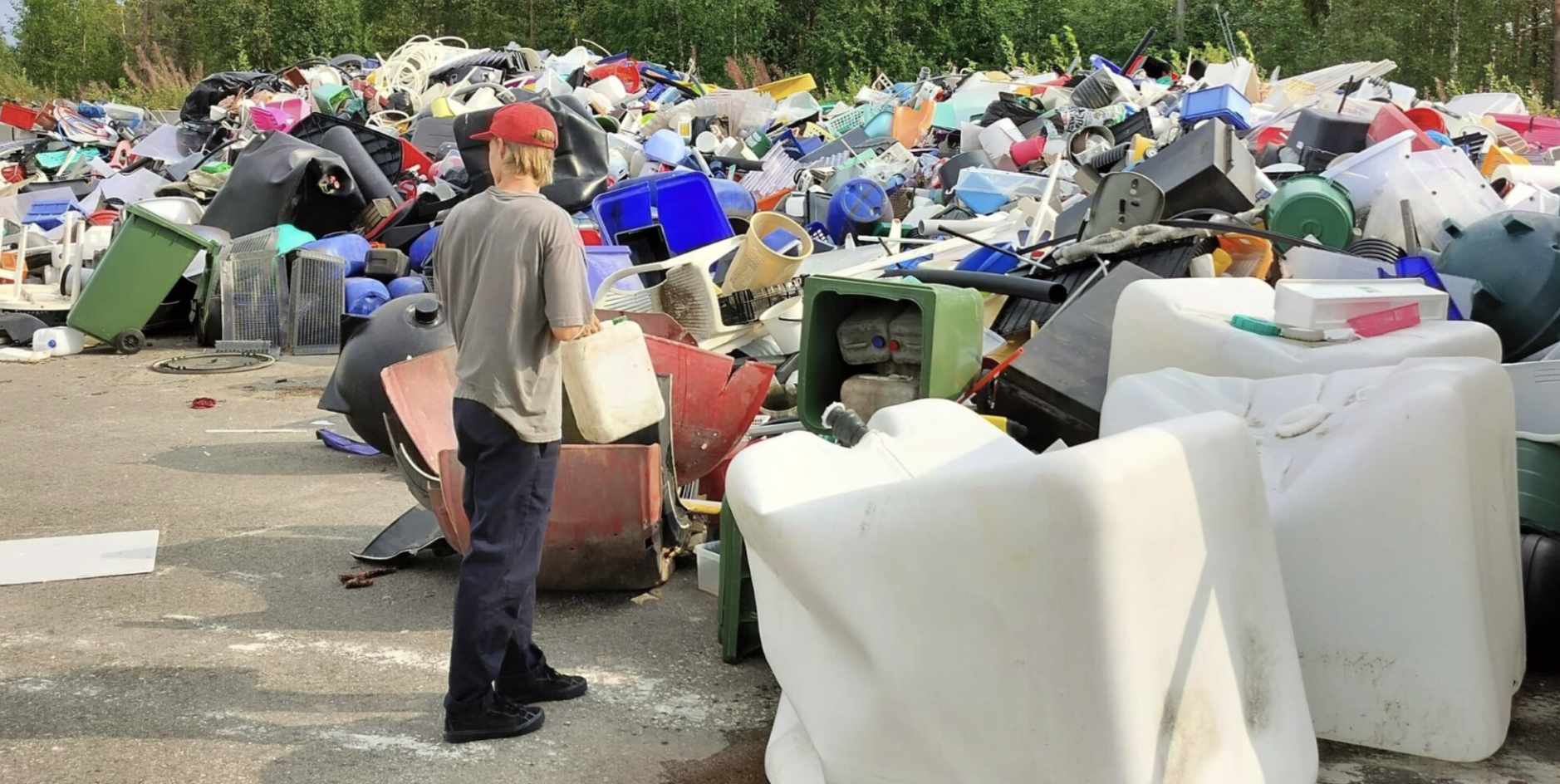}
    \caption{The existing area for plastics collection at the recycling center.}
    \label{fig:residum_real}
\end{figure}

\begin{figure}
    \centering
    \includegraphics[width=1\linewidth]{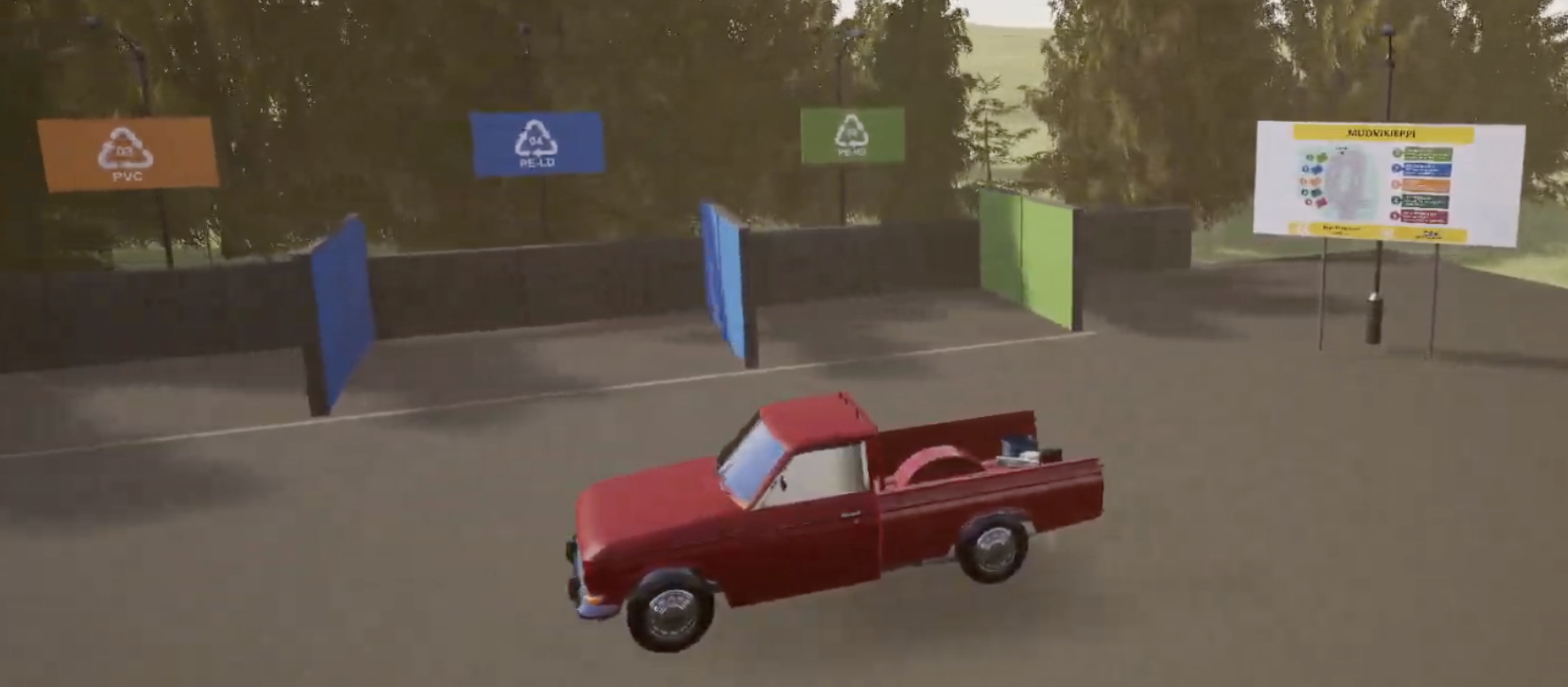}
    \caption{VR model of the planned recycling area.}
    \label{fig:residum_vr}
\end{figure}

\begin{figure}
    \centering
    \includegraphics[width=1\linewidth]{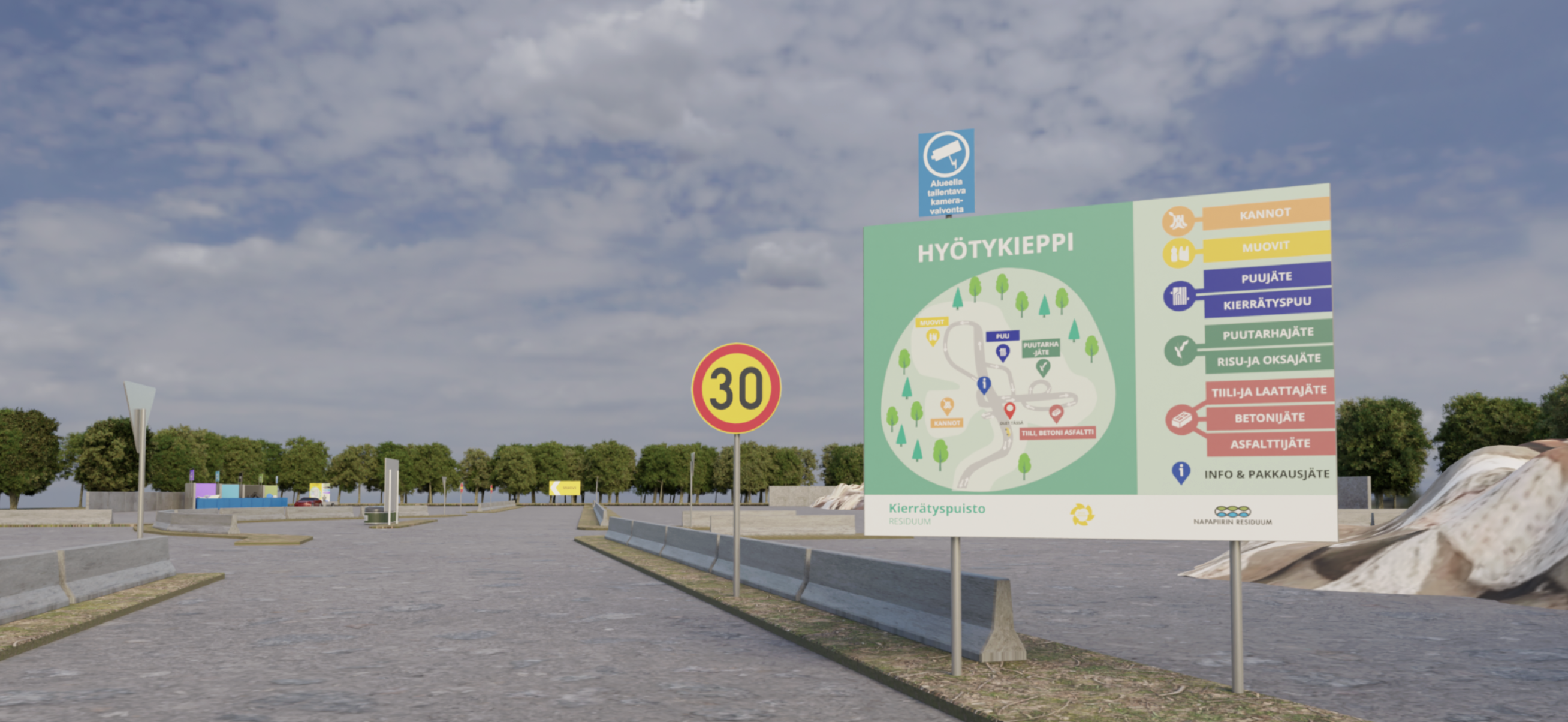}
    \caption{VR model of the entrance to the recycling center. The planned plastic recycling area is marked in yellow on the map and is visible in the center-left of the image. }
    \label{fig:residum_vr_entrance}
\end{figure}

The second case focuses on a public recycling center that was being planned and built by a regional waste management company \cite{etto2024designing}. The company sought new ways to communicate the sorting rules for plastics and to collect feedback on the planned layout and signage before the center was constructed. Plastic packaging forms a significant part of household waste, and incorrect sorting leads to contamination and losses in recycling streams \cite{vogt2021recycling}. Service design methods were already used for mapping customer paths and touchpoints, and the project team wanted to explore VR as an extra tool for representing the future site and involving stakeholders in its design.

\subsection{VR prototype and interaction design}

To create the VR simulation, the design team first visited the planned recycling area and collected photographs, videos and layout drawings (Figure \ref{fig:residum_real}). Based on these materials, an initial 3D model of the site was created in Blender, including roads, waste containers, buildings and other structures (Figures \ref{fig:residum_vr} \& \ref{fig:residum_vr_entrance}). This model was then imported into the Unreal Engine 5 game engine, where interaction logic and environmental details were added. Ambient sound, such as bird song, contributed to a recognisable outdoor atmosphere, and lighting settings were adjusted to support visibility and comfort in the headset.

The VR application ran on a head-mounted display with handheld controllers. Player movement utilised the controller thumbstick for short teleport jumps, providing users a way to explore the yard and reposition themselves near containers (Figure \ref{fig:residum_vr_entrance}). The main interactive elements were plastic items that could be grabbed and carried, and the recycling containers that functioned as targets for sorting. When a player released an item near a container, the system checked whether the choice matched the waste type and provided immediate feedback through audio and visual cues.

The simulation included three levels of signage that mirrored the planned physical information design at the site. At a glanceable level, simple icons and colour coding indicated main fractions and directions. A second level presented short text labels and category names on signs near each container group. A third level offered more detailed instructions, accessed when the player approached information boards or turned toward specific panels. This layered structure followed service design practice for complex environments and aligned with prior work on visualising service paths in spatial settings \cite{bae2014visual,plotzky2021virtual,kaminska2017virtual}.

\begin{figure}
    \centering
    \includegraphics[width=1\linewidth]{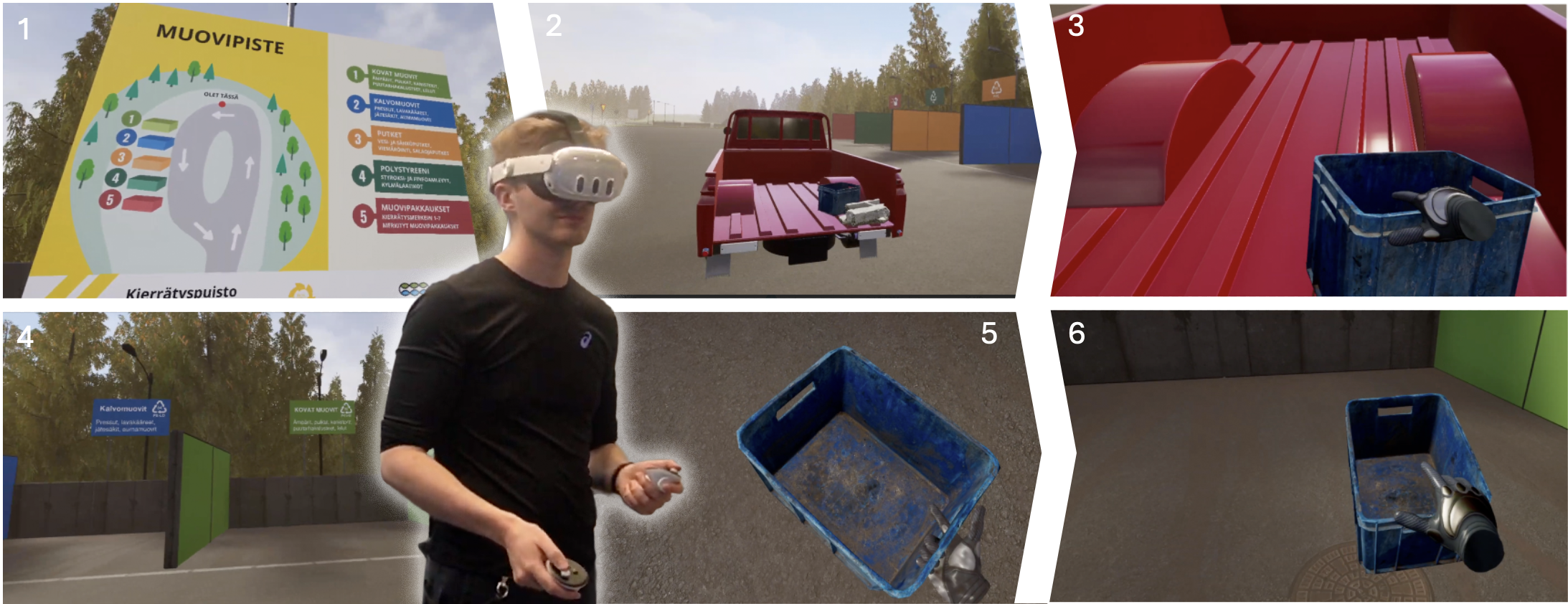}
    \caption{User interaction in the VR recycling center: 1) Viewing the map, 2) Parking next to the plastics area, 3) Taking a plastics item. 4 Identifying the correct recycling container for the plastic type. 5-6) carrying the object and depositing it in the recycling container.}
    \label{fig:residium_user_interaction}
\end{figure}

\subsection{Use in service design activities}

The VR prototype was used in three workshops that covered a complete service design process for the recycling center. These workshops differed in purpose, participants and the maturity of the VR material:

\begin{itemize}
    \item \textbf{Workshop 1 (early concept).}  
    Five company employees explored 360° site images and an early 3D model. VR served mainly as contextual material that supported discussions about traffic flow, container placement and information needs for plastics and other fractions.

    \item \textbf{Workshop 2 (co-design and evaluation).}  
    Thirteen participants, including both company staff and external stakeholders, engaged with a more complete interactive VR prototype. Participants took turns navigating the yard, locating the plastics area and testing sorting tasks. The group discussed layout, signage clarity and user experience while annotating printed maps and photographs.

    \item \textbf{Workshop 3 (refinement and final review).} 
    Six company employees reviewed an updated VR simulation aligned with the near-final design of the physical center. VR enabled scenario walkthroughs, where participants rehearsed typical visitor journeys and refined signage, traffic rules and plastics-related touchpoints.
\end{itemize}

\subsection{Findings}

Across the workshops, participants reported that the VR simulation helped them picture the future recycling center more concretely than static drawings alone, in line with earlier reports on VR prototypes in service design \cite{boletsis2018virtual,bae2014visual}. The ability to move at ground level, look around freely and handle virtual items supported discussion about sight lines, wayfinding and the clarity of signage. Company employees highlighted that the VR view made gaps and inconsistencies in information placement more visible, e.g., when a plastic container was partially hidden by other structures or when category labels looked too similar from a distance.

The case highlighted several practical and social challenges. Wearing a headset limits eye contact and non-verbal cues, which can hinder group discussion in co-located workshops \cite{colley2020exploring,gugenheimer2018demonstration}. Designers and participants noted that only one person could be fully immersed at a time, which created a division between the person in VR and the rest of the group. Large-screen projection of the headset view partly mitigated this issue by giving others a window into the virtual scene, yet the sense of a shared experience remained weaker than around physical prototypes on a table.

From the design team's perspective, VR brought benefits for remote collaboration as well. The same virtual environment could be used in distributed meetings, where stakeholders connected from different locations and explored the site together or sequentially \cite{makela2020virtual,rivu2021remote,lee2022hci}. This supported project progress in a context where long travel distances and seasonal conditions make frequent site visits difficult. At the same time, the effort required to create and update the VR model raised questions about trade-offs between fidelity and flexibility, echoing broader discussions in service prototyping research \cite{razek2019comparing,schneider2011service}.

\subsection{Case-specific reflections for service design}

The recycling center simulation illustrates how VR can function as a service design tool that links spatial layout, information design and customer experience in a single artefact. In early phases, simple 3D models and 360° imagery can support joint sensemaking about context and constraints. In later phases, an interactive prototype gives stakeholders a way to rehearse customer paths, identify pain points and test revised solutions before construction or large-scale communication investments.

The case underlines both opportunities and limitations of VR in collaborative design. On the opportunity side, immersion and spatial fidelity give stakeholders a strong feel for distances, visibility of signage and the physical effort required to move waste at the site, which is important for inclusive plastics recycling services. On the limitation side, lack of shared embodiment and limited multi-user tools make it harder to run fully collaborative ideation in VR, which connects to concerns in participatory design about balancing digital tools and face-to-face interaction \cite{holmlid2012participative,juaneda2020unified}. Designers in this project suggested combining VR sessions with conventional workshop practices such as group sketching and physical mock-ups, instead of seeing VR as a replacement.

Overall, the case shows that a VR simulation of a plastics recycling center can act both as an educational experience for future users and as a central artefact in a structured service design framework.

\section{Discussion}

This section brings the two cases together and considers how VR tools contributed to education and service design of plastic recycling. The aim is to identify the roles played by VR, the practical choices involved and the way the two cases point toward a possible process pattern for VR use in service design work.

\subsection{Positions of VR along the recycling journey}

The two cases support different points in the recycling service journey. The Clean Cabin Escape introduces sorting activities in a home-like setting. Players handle familiar objects, apply local rules and receive feedback that encourages repeated attempts. Prior work on recycling education and game-based learning highlights how repetition and immediate cues can support learning \cite{bennett2021informing,Sidian2024-dy,Novo2024-fx,Bai2024-nc}.

The recycling center case focuses on a later stage in the recycling journey where users encounter signage, routes, containers and practical tasks. VR allowed participants to move through the planned space, assess clarity and identify possible points of confusion. Earlier studies on servicescapes and customer paths describe how layout and visibility influence experience \cite{bitner1992servicescapes,ro2017making}. Staff and designers were able to use the virtual scene to rehearse specific actions and refine details before physical site construction.

\subsection{VR within a wider service design toolset}

Both projects were part of broader design work that used sketches, maps and workshop activities. VR acted as one material within that mix. This aligns with descriptions of service design as a process that combines visualisation, enactment and discussion \cite{miettinen2017introduction,stickdorn2018service}.
In the cabin case, VR can be used together with lessons, printed guides and other educational tasks. Observations from its use can highlight which plastics items participants find most uncertain and inform updates to communication materials \cite{lidia2018encourage,palmquist2022design}.
In the recycling center case, VR complemented the group work where participants annotated printed drawings, proposed changes and reflected on visibility and flow. VR provided a shared view, while sketches supported quick revisions. Earlier work notes similar patterns where VR is combined with familiar design artefacts \cite{boletsis2018virtual,colley2020exploring}.

\subsection{Design choices and trade offs with VR use}

The cases point to several practical considerations. Access and comfort influence how groups can work with VR. New users may need clear guidance. In group settings, only one person is fully inside the headset at a time, which places importance on the projected view for observers \cite{gugenheimer2018demonstration}. Fidelity influences what participants can evaluate. The center scene aimed to resemble the planned site closely, which supported assessment of visibility and layout but required significant modelling. Studies on VR for public space and exhibition design report similar observations \cite{pierroux2021exhibitions,hafner2016virtual}. The cabin game required a balance between simplicity and item variety. Feedback design shapes experience. The cabin game used clear audio and visual cues to support exploration. The center scene relied on subtler cues since it aimed to resemble a visit to a real facility \cite{Novo2024-fx,venturi2025trash}.

The two cases also differ in their social configuration. The cabin experience suits individual practice. The center simulation aimed to support group discussion, where one user acted in VR while others observed. Earlier studies of VR fieldwork note the importance of providing a mirrored display of the VR view to the group and verbal coordination in such settings \cite{makela2020virtual,rivu2021remote}.

\subsection{A process pattern suggested by the two cases}

Across both cases, a similar sequence appeared in how VR material was introduced and used. This sequence may be useful for other projects that consider VR as part of service design work.

\begin{itemize}
    \item \textbf{Early concept.}  
    VR material: 360° images and simple spatial layouts.  
    Purpose: help participants form a shared view of the context and main features of the site or setting.

    \item \textbf{Co-design.}  
    VR material: an interactive prototype that supports movement, inspection and basic tasks.  
    Purpose: explore signage, flow, navigation and plastics-related decisions with stakeholders.

    \item \textbf{Refinement.}  
    VR material: an updated interactive scene that reflects near-final layout choices.  
    Purpose: review details such as visibility, placement and instructions before implementation.
\end{itemize}

This sequence does not replace the broader set of service design methods. It simply reflects the way VR contributed in these two cases and may be applicable in other contexts where teams combine VR with workshops, sketches and physical artefacts.

\subsection{Implications for plastics recycling and circular economy services}

Based on the two presented case studies, VR can support plastics recycling by presenting sorting tasks and facility layouts in concrete scenes that both educate individuals and provide project teams with materials to discuss practical decisions within broader design work. There are opportunities for further work. Longer term studies could examine whether VR-based learning affects actual sorting behaviour, e.g. exploring how VR combines with mobile apps or interactive installations. Service design methodology research could refine how teams integrate VR scenes with workshop activities and other tools in different service settings.

\section{Conclusion}

This paper has examined how VR tools can support plastics recycling through two application cases positioned at different points in the recycling journey. the Clean Cabin Escape presents a domestic-scale VR escape room that allows players to practise sorting tasks in a focused setting. The VR simulation of a plastics recycling center offers a way to explore a planned site before construction, helping participants comment on layout, signage and practical aspects of the visit.

Viewed together, the cases show that VR can make both household sorting and larger recycling facilities easier for people to discuss. VR scenes can draw attention to features of recycling that are difficult to convey through static material and can prompt reflection on how citizens carry out sorting tasks in everyday life or navigate public recycling sites. The projects also highlight practical issues such as access to headsets, onboarding for new users, the work required to create and maintain VR scenes and the need to match content with local recycling rules. The cases indicate that VR is most useful when directed at clear questions within wider design efforts. VR can sit alongside workshops, physical artefacts and communication material, offering concrete situations for practice, review or group discussion. Its contribution lies in presenting tasks and scenarios in a way that supports both citizen learning and collaboration among stakeholders.

\begin{acks}
This study was funded by the Rethinking Plastics in a Sustainable Circular Economy (PlastLIFE) project (LIFE21-IPE-FI-PlastLIFE). The PlastLIFE project was co-funded by the European Union. Views and opinions expressed are however those of the authors only and do not necessarily reflect those of the European Union or CINEA. Neither the European Union nor the granting authority can be held responsible for them.
\end{acks}

\bibliographystyle{ACM-Reference-Format}
\bibliography{11-references}

\end{document}